\documentclass[twocolumn]{aastex631}
\usepackage{svg}

\usepackage{CJKutf8}
\usepackage{color}

\usepackage{mathrsfs}
\usepackage{rsfso}
\usepackage{rotating}

\usepackage{graphicx}	
\usepackage{amsmath}	
\usepackage{amssymb}	

\newcommand\lsim{\mathrel{\rlap{\lower4pt\hbox{\hskip1pt$\sim$}}
\raise1pt\hbox{$<$}}}
\newcommand\gsim{\mathrel{\rlap{\lower4pt\hbox{\hskip1pt$\sim$}}
\raise1pt\hbox{$>$}}}

\shortauthors{Faridani et al.}

\graphicspath{{./}{figures/}}

\defcitealias{Morton+16}{Mo16}
\defcitealias{Valizadegan+22}{V22}
\defcitealias{Mayo+18}{Ma18}

\shorttitle{More Likely Than You Think}
\shortauthors{Faridani et al.}

\graphicspath{{./}{figures/}}

\begin{document}

\title{Peekaboo: Secular Resonances from Evolving Stellar Oblateness Impede Transit Detections

}

\author[0000-0003-3799-3635]{Thea H. Faridani}
\correspondingauthor{Thea H. Faridani}
\email{thfaridani@astro.ucla.edu}
\affiliation{Department of Physics and Astronomy, University of California, Los Angeles, CA 90095, USA}
\affiliation{Mani L. Bhaumik Institute for Theoretical Physics, Department of Physics and Astronomy, UCLA, Los Angeles, CA 90095, USA}

\author[0000-0002-9802-9279]{Smadar Naoz}
\affiliation{Department of Physics and Astronomy, University of California, Los Angeles, CA 90095, USA}
\affiliation{Mani L. Bhaumik Institute for Theoretical Physics, Department of Physics and Astronomy, UCLA, Los Angeles, CA 90095, USA}

\author[0000-0001-8308-0808]{Gongjie Li}
\affiliation{School of Physics, Georgia Institute of Technology, Atlanta, GA 30332, USA}

\author[0000-0002-7670-670X]{Malena Rice}
\affiliation{Department of Astronomy, Yale University, New Haven, CT 06520, USA}

\author[0000-0001-8342-7736]{Jack Lubin}
\affiliation{Department of Physics and Astronomy, University of California, Los Angeles, CA 90095, USA}

\begin{abstract}

Secular resonances in exoplanet systems occur when two or more planets have commensurabilities in the precession rates of their orbital elements, causing an exchange of angular momentum between them. The stellar gravitational quadrupole moment, which evolves over time due to stellar spin-down over the first $\sim 100\,$Myr, causes these resonances to sweep through the parameter space (of masses and semimajor-axis ratios), affecting a wider variety of systems than when spin-down is neglected. The angular momentum exchange in these resonances typically aligns the outer planets' orbits together while misaligning the innermost planet from its companions. Here, we explore how resonance-induced (mis-)alignments between planets affect the transit outcome. We use the three-planet Kepler-619 system as a concrete case study that is relatively likely (approximately $1$:$1$ odds) to have undergone a sweeping resonance. Consequently, if the outer planets are initially misaligned, a sweeping-resonance-induced push toward alignment in the outer planets' orbits substantially increases the probability that they are observed transiting together. However, the resulting misalignment of the innermost planet from its outer companions, reduces the overall probability that Kepler-619 is observed as a multi-planet transiting system when considering any pair of planets or all three together. This implies that sweeping resonances (which may occur in $20\%$ of systems) have a deleterious effect on the transiting exoplanet sample, suppressing the detection of multi-planet transiting systems in favor of apparent single-planet systems.

\end{abstract}

\section{Introduction}

After contracting from their pre-main sequence phase, sun-like stars about $100\,$ Myr of age have spin periods ranging from about $\sim 0.5$ $-$ $10$ days \citep[see e.g.,][]{Gallet+13}. At this age, these stars experience a rapid deceleration of their spins due to magnetic braking--the torquing of the wind by the star's magnetic field transferring angular momentum away from the star \citep[e.g.,][] {Parker+58,Schatzman+62,Kraft+67,WeberDavis+67,Mestel+68,Skumanich+72}. When sun-like stars age to $600-1000\,$Myr, the correlation between the strength of magnetic braking and the stellar spin rate brings all sun-like stars to similar spins.

For planetary systems around sun-like stars, the host stars' spin rates at early times have the potential to significantly alter their planets' orbital angles. Stars' spins rotationally flatten them at their poles, altering their gravitational potential. This deviation from a point-mass potential is, to leading order, parametrized by the gravitational quadrupole moment, $J_2$. The effect of $J_2$ is to induce a precession in the arguments of periapsis and longitudes of ascending node of all orbiting bodies \citep[see e.g.,][]{Murray+00book}. These precessions can have a wide range of effects, such as misaligning tightly-packed systems \citep[e.g.,][]{Spalding+16,Becker+20,Brefka+21,chen+22,Faridani+23,Faridani+24} or aiding tidal migration \citep[e.g.,][]{Pu+19,Millholland+20}. As spin rates decrease over time, $J_2$ does as well, reducing the rates of these precessions.

Mutual gravitational interactions between the planets also cause precessions in these angles at a wide range of frequencies. If there are first-order commensurabilities between the slower of these frequencies (i.e., frequencies with periods much longer than the orbital periods in the system), the orbiting bodies enter what is known as a ``secular resonance.'' In such a resonance, the involved bodies rapidly exchange angular momentum \citep[see e.g.,][]{Murray+00book}. The results of this exchange can range from planetary interactions with tides \citep[see e.g.,][]{Hansen+15}, to circularize and migrate \citep[potentially causing chaotic evolution through overlap of many higher-order resonances, see e.g.,][]{LithwickWu+11,Petrovich+19,O'Connor+22}, to coupling with planetary spin and tilting planets' spin axes \citep[see e.g.,][]{Millholland+19}. 

The evolving stellar $J_2$ early in the star's life can significantly affect when and where secular resonances occur in an exoplanetary system \citep[e.g.,][]{Brefka+21,Faridani+23,Faridani+24}. The decreasing $J_2$ can move the locations (in semimajor-axis ratio space) of secular resonances, dramatically increasing the range of planetary system architectures where resonances can occur \citep{Faridani+24}. Specifically, \citet{Faridani+24} showed that out of a sample of $30$ observed, three-planet systems orbiting stars below $6200\,$K\footnote{$6200\,$K denotes the Kraft break, above which stars are not expected to form the convective envelopes necessary to produce a strong magnetic field and efficiently undergo magnetic braking. Stars above $6200\,$K therefore retain high spins for much longer \citep[see e.g.,][]{Kraft+67}. Note, however, that  \citet{Beyer+24} find that the Kraft Break -- defined specifically as a rotational discontinuity, rather than a dividing line of stellar structure -- is centered at $\sim6550\,$K. In this work, we do not discuss any star hotter than $6200\,$K, but note that stars between $6200\,$-$6550\,$K may also spin down and cause resonance-induced misalignments in their planets.}, that $6$ of them (approximately $20\%$) exist in configurations where a sufficiently high initial stellar $J_2$ would cause the planets, at some point during the star's spin-down, to enter secular resonance and exchange angular momentum. This exchange typically misaligns the inner planet from the outer two, which align with each other in the process \citep[e.g.,][]{Faridani+24}. Therefore, systems susceptible to resonances mediated by evolving $J_2$ may experience significant orbital changes depending on whether or not the initial stellar spin was large enough to reach the critical $J_2$ required for resonance. In this work, we calculate how this (mis-) aligned configuration affects transit probability.

As a test case, we consider the Kepler-619 system. Kepler-619 is a sun-like star with mass $1.09\,M_\odot$ and radius $1.11 R_\odot$, hosting three planets (`c', `b', and `d') with periods of $1.21\,d$, $5.4\,d$, $11.68\,d$ and radii of $1.69\,R_\oplus$, $3.16\,R_\oplus$, and $4.38\,R_\oplus$, respectively \citep[][]{Morton+16,Valizadegan+23}.

This paper is structured as follows: in Section \ref{sec:math}, we introduce the mathematical model. In Section \ref{sec:obs} we explore the effects of sweeping $J_2$ resonances on the transit probabilities for observed three-planet system Kepler-619. 
In Section \ref{sec:conclusion} we discuss and conclude.

\section{Sweeping Secular Resonances and the Relation to $J_2$}\label{sec:math}

\begin{figure*}
\centering
    \includegraphics[width=\linewidth]{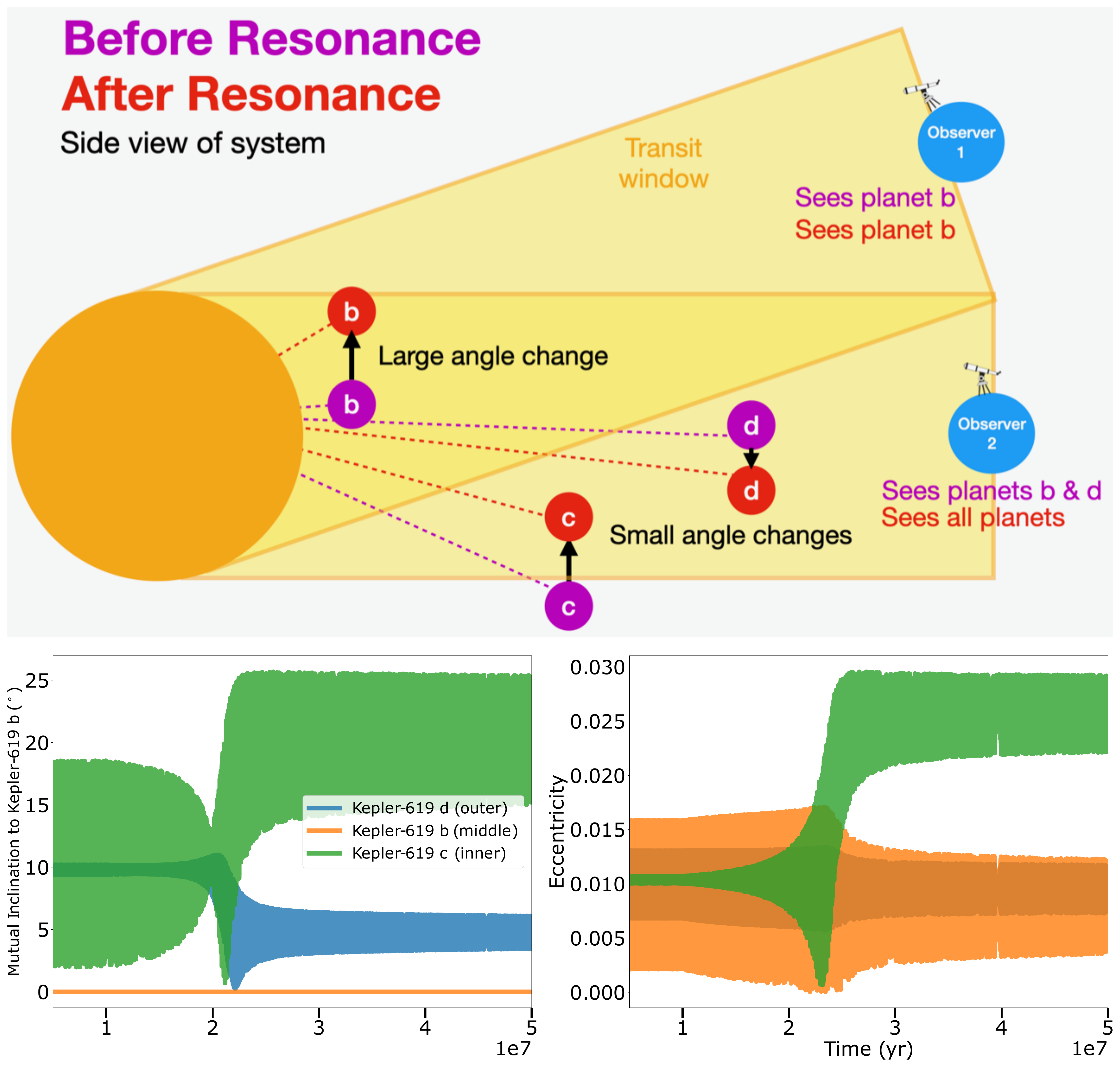}
    \caption{
    (Top) Qualitative schematic of impact-parameter geometry before and after sweeping resonance. Angles, distances and radii are not to scale. \\  
    (Bottom) Time evolution of the inclinations of the Kepler-619 system with high initial $J_2$. The inner, middle, and outer planets were initialized with inclinations of $10.3^\circ$, $6.1^\circ$, $4.5^\circ$, eccentricities of $0.010$, $0.006$, and $0.0125$, and longitudes of ascending node of $1.8$, $3.1$, $4.9$ radians, respectively. These initial conditions were chosen from the Monte Carlo described in Section \ref{sec:obs}, and therefore, were chosen arbitrarily to demonstrate clearly what resonance-induced misalignments look like as functions of time. 
    }
    \label{fig:TimeEvolution}
\end{figure*}

The gravitational influence of each planet in a multi-planet system on its neighbors over secular timescales is described by the Laplace-Lagrange secular model \citep[see e.g.,][]{Murray+00book}. The precessions of inclination and longitude of ascending node (eccentricity and longitude of periapsis) are determined by the Laplace-Lagrange formalism, defined, for a host star of mass $M_\star$, radius $R_\star$, and quadrupole moment $J_2$, by
\begin{equation}\label{eq:R}
\begin{aligned}
\mathcal{R}_j^{(\rm sec)}=& n_j a_j^2\bigg[\frac{1}{2} A_{j j} e_j^2+\frac{1}{2} B_{j j} I_j^2  \\  
&+\sum_{k=1, k \neq j}^N A_{j k} e_j e_k \cos \left(\varpi_j-\varpi_k\right) \\ 
&+\sum_{k=1, k \neq j}^N B_{j k} I_j I_k \cos \left(\Omega_j-\Omega_k\right) \bigg] ,
\end{aligned}    
\end{equation}
\begin{equation}
    \begin{aligned}
    A_{jj} =& n_j\bigg[ \frac{3}{2}J_2 \left( \frac{R_\star}{a_j} \right)^2 \\
    &+ \sum_{k=1,k\neq j} \frac{m_k}{4\pi \left( M_\star +m_j\right)} \alpha_{jk} \overline{\alpha}_{jk} f_\psi (\alpha_{jk}) \bigg] \ ,        
    \end{aligned}
    \label{eq:A_matrix_start}
\end{equation}

\begin{equation}
    A_{jk} = -\frac{n_j m_k}{4\pi \left( M_\star +m_j\right)} \alpha_{jk} \overline{\alpha}_{jk} f_{2\psi} (\alpha_{jk})  \ ,
\end{equation}
\begin{equation}
\begin{aligned}
    B_{jj} &= -n_j\bigg[ \frac{3}{2}J_2 \left( \frac{R_\star}{a_j} \right)^2 \\
    &+ \sum_{k=1,k\neq j} \frac{ m_k}{4\pi \left( M_\star +m_j\right)} \alpha_{jk} \overline{\alpha}_{jk} f_\psi (\alpha_{jk}) \bigg] \ ,
    \end{aligned}
\end{equation}\label{eq:bmat}
\begin{equation}
    B_{jk} = \frac{n_j m_k}{4\pi \left( M_\star +m_j\right)} \alpha_{jk} \overline{\alpha}_{jk} f_\psi (\alpha_{jk})  \ ,
    \label{eq:B_matrix}
\end{equation}
where
\begin{equation}
    \alpha_{jk} = \min \left(\frac{a_j}{a_k},\frac{a_k}{a_j}  \right) \ ,
\end{equation}
\begin{equation}
    \overline{\alpha}_{jk} = \min \left(\frac{a_j}{a_k},1  \right) \ ,
\end{equation}
\begin{equation}
    f_{\psi}(\alpha_{jk})=\int_{0}^{2 \pi} \frac{\cos \psi}{\left(1-2\alpha_{jk} \cos \psi+\alpha_{jk}^{2}\right)^{2}} \mathrm{d} \psi\ ,
\end{equation}
and
\begin{equation}
    f_{2 \psi}(\alpha_{jk})=\int_{0}^{2 \pi} \frac{\cos 2 \psi}{\left(1-2\alpha_{jk} \cos \psi+\alpha_{jk}^{2}\right)^{\frac{3}{2}}} \mathrm{d} \psi \ ,
    \label{eq:fpsi_end}
\end{equation}
\begin{equation}
    J_2 = \frac{1}{3} k_2 \left(\frac{S}{S_b}\right)^2,
    \label{eq:j2_formula}
\end{equation}
\begin{equation}\label{eq:LL_orbital_elements}
\begin{aligned}
\dot{e}_j & =-\frac{1}{n_j a_j^2 e_j} \frac{\partial \mathcal{R}_j^{(\rm sec)}}{\partial \varpi_j}, & \dot{\varpi}_j=+\frac{1}{n_j a_j^2 e_j} \frac{\partial \mathcal{R}_j^{(\rm sec)}}{\partial e_j}, \\
\dot{I}_j & =-\frac{1}{n_j a_j^2 I_j} \frac{\partial \mathcal{R}_j^{(\rm sec)}}{\partial \Omega_j}, & \dot{\Omega}_j=+\frac{1}{n_j a_j^2 I_j} \frac{\partial \mathcal{R}_j^{(\rm sec)}}{\partial I_j} ,
\end{aligned}    
\end{equation}

\begin{equation}\label{eq:GR_orbital_elements}
    \dot{\omega}_{j,GR} = \frac{3 (GM_\star)^{3/2}}{c^2 a_j^{5/2} (1-e_j^2)} \ .
\end{equation}
where the $j$th planet has a mass $m_j$,  semi-major axis $a_j$, mean motion $n_j$), eccentricity $e_j$, longitude of ascending node $\Omega_j$, longitude of periapsis $\varpi_j$, and inclination with respect to the star's spin axis $I_j$. $S$ corresponds to the stellar spin, and $S_b$ to the breakup spin defined by $S_b = \sqrt{GM_\star/R_\star^3}$. Integrating these equations provides the evolution of all the orbits in the system. Unless specified otherwise, all simulations in this work are calculated using numerical integrations of Equations \ref{eq:LL_orbital_elements} and \ref{eq:GR_orbital_elements}.

If the proper precession of an additional orbiting body's longitude of ascending node or longitude of periapsis matches an eigenfrequency of the B or A matrix, respectively, then that body has entered a linear secular resonance. In such a resonance, angular momentum is exchanged between bodies at an enhanced rate, potentially destabilizing the system or significantly altering its architecture. As can be seen by Equations (\ref{eq:A_matrix_start}) and (\ref{eq:B_matrix}), a change in the host star's $J_2$ as it spins down will cause a change in the eigenfrequencies of the system. This results in a change over time of the range of orbits that are in secular resonance. This is generically called a ``sweeping resonance'', and is discussed thoroughly in the context of $J_2$ evolution in \citet{Faridani+24}. 

We illustrate the impact of sweeping resonances on Kepler-619, caused by evolving $J_2$, in the bottom panels of Figure \ref{fig:TimeEvolution}. Stellar $J_2$ is initialized to a high value of $6\times 10^{-4}$. This corresponds to a spin period of 1.6 days, approximately at the $75$th percentile for young sun-like stars \citep[see e.g.,][]{Gallet+13}. This is higher than the critical value of $J_2 \sim 10^{-4}$ (corresponding to a $4$ day spin period, approximately $40$th percentile for initial spin) where the system enters the secular resonance. At $t\sim 2\times 10^7$ years into the simulation, $J_2$ hits the critical value, which brings the dominant eigenmode of the inner planet's longitude of ascending node precession equal to the dominant mode of the outer planets' precession, entering the resonance. The mutual inclination of the inner planet with respect to the outer two is then transitioned to a higher value, and, to conserve angular momentum, the mutual inclination of the outer two planets is reduced significantly. This resonance-induced alignment occurs over approximately $5\times 10^6$ years, many times longer than the period of the inner planet's dominant eigenmode. In order for these changes in alignment to occur, the outer planets must have some initial mutual inclination; otherwise, if all planets are aligned, they cannot torque one another out of alignment. 

The effect of the resonance-induced alignments on the detectability of a three-planet system via transit is stark, as will be quantified in the following sections. Any reduction in the mutual inclination between the outer planets will increase the probability that they can be observed transiting together, and the misalignment of the inner planet will reduce the probability that it transits together with one of its companions. These contrasting effects mean that the change in probability of detecting all three planets transiting together is not a priori clear. However, what is certain is that the alignment of the outer planets will increase their pairwise co-transit probability, and the misalignment of the inner planet to its companions will decrease its pairwise co-transit probabilities.

\section{Test Case of Observed Transiting Systems}\label{sec:obs} 

\subsection{Monte Carlo Design}

To investigate the impact of the misalignments caused by sweeping secular resonances from evolving $J_2$, we explore how ranges of initial conditions impact the system. For each selection of initial conditions, we integrate numerically Equations (\ref{eq:LL_orbital_elements}) with the addition of a GR correction to $\dot{\varpi}$ described in Equation \ref{eq:GR_orbital_elements}. Inclinations were drawn from Rayleigh distributions with various $\sigma_i$ values, ranging linearly from $0.01\,{\rm rad} \approx 0.6^\circ$ to $0.15\,{\rm rad} \approx 8.6^\circ$. The end of this range was chosen to produce mutual inclinations much larger than the observed median mutual inclination distribution between transiting planets in three-planet transiting systems, so as to cover a large range of potential distributions \citep[see][who estimate the median mutual inclination between planets in a three-planet transiting system is $\sim 4.8^\circ$]{Zhu+18Architecture}.  The Rayleigh distribution is defined by 
\begin{equation}
p(x) = \frac{x}{\sigma^2} e^{-x^2 /\left(2 \sigma^2\right)},
\end{equation} 
and is commonly used to model exoplanet inclinations \citep[see e.g.,][]{Fabrycky+14}. Planetary eccentricities were drawn from radius-dependent beta distributions, whose $\alpha$ and $\beta$ parameters were taken from Table S2 of \citet{Gilbert+25}, where we adopt $R_{\rm gap} = 1.8 R_\oplus$. Initial $J_2$ values were drawn uniformly from $10$ bins spaced equally in log between $2\times 10^{-5}$ and $6\times 10^{-4}$ inclusive. In total, we run 22,000 runs up to $5\times 10^7$~yrs, which is well beyond the time the resonance takes place \citep{Faridani+24}.

To calculate the transit probabilities of the planets, we use the open-source CORBITS package, which calculates single- and multiple-transit probabilities given a system's orbital elements \citep[][]{Brakensiek+16CORBITS}\footnote{We used a fork that was wrapped in Python by Samuel Yee \url{https://github.com/samuelyeewl/CORBITS}}. We calculate the mean probabilities at the end of the simulation using the orbital elements at 1000 times spaced evenly over the last 1\% of the simulation.

\begin{figure}
\centering
    \includegraphics[width=\linewidth]{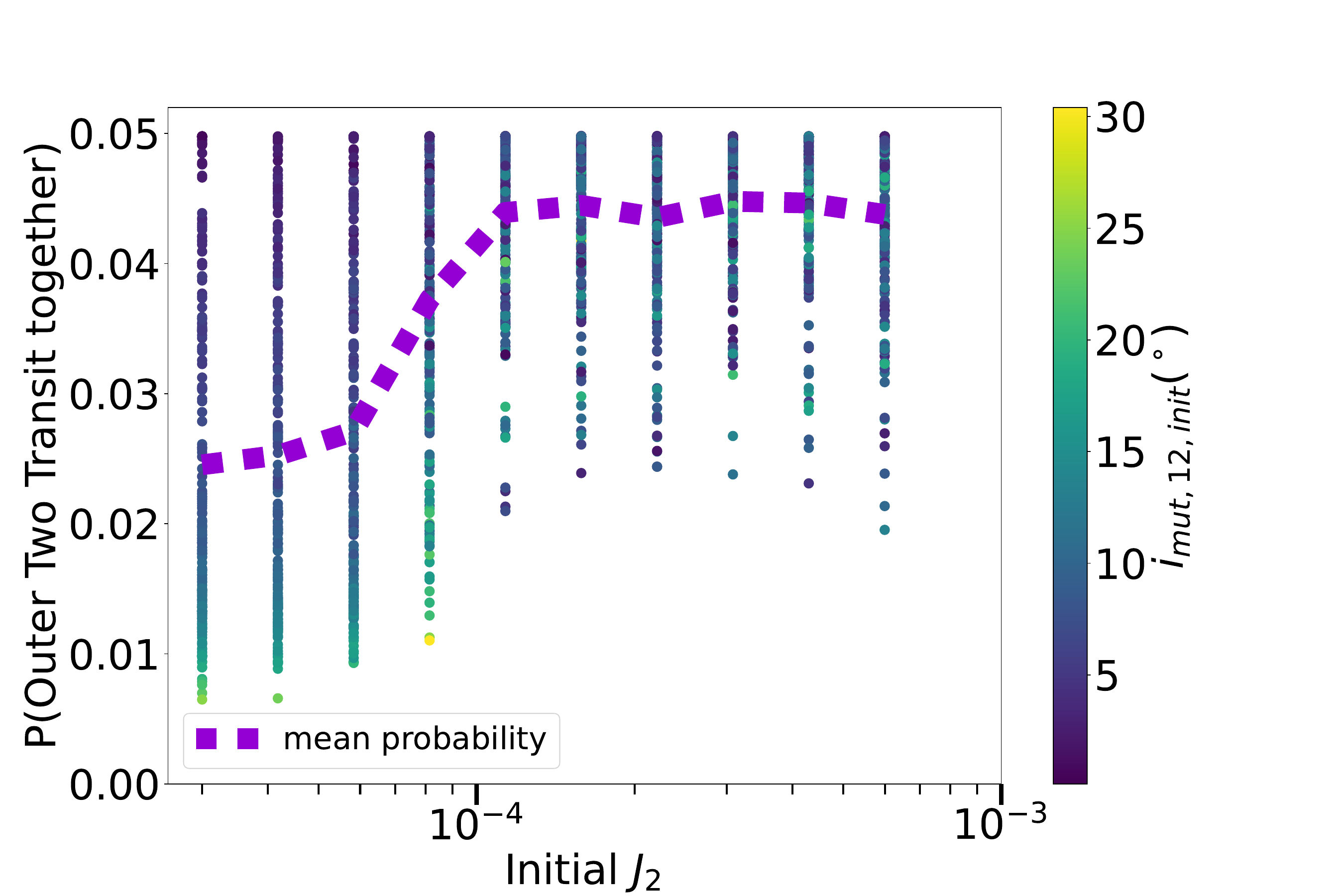}
    \caption{
    Overview of the effects of $J_2$ on transit probability for the outer planets in Kepler-619. The probability that the outer two observed planets transit together  when viewed from a random angle (irrespective of whether the inner planet transits at this angle) is shown for 2000 integrations of orbital parameters as a function of initial stellar $J_2$. The inclination distribution for these integrations used a Rayleigh-sigma of $\sigma_i = 0.08 \,{\rm rad} \approx 4.8^\circ $. The purple curve represents the mean probability as a function of $J_2$.
    }
    \label{fig:K619OuterPlanets}
\end{figure}

\begin{figure}
\centering
    \includegraphics[width=\linewidth]{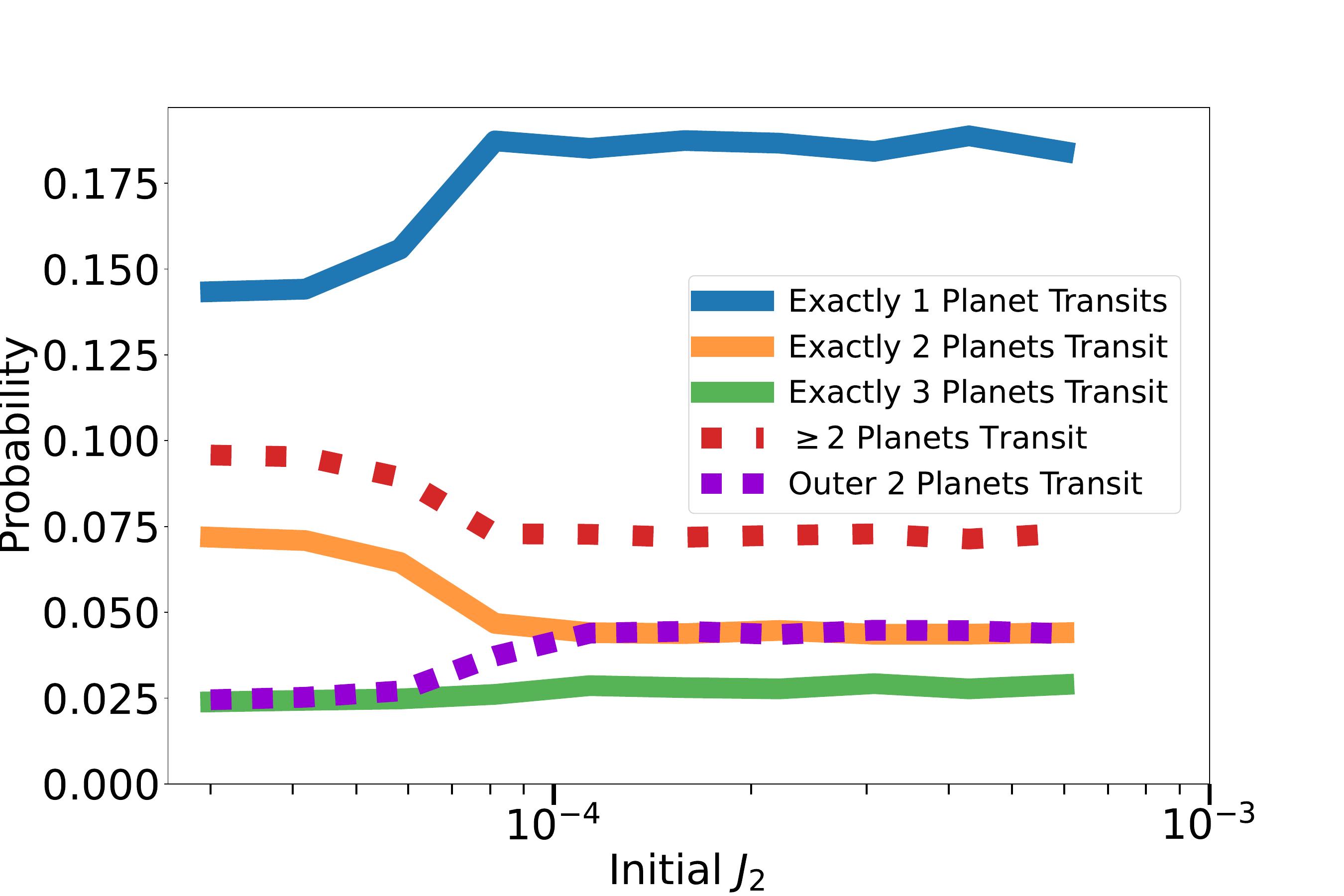}
    \caption{
    Overview of the effects of $J_2$ on average transit probability for the planets in Kepler-619 over a Monte Carlo of 2000. Longitudes of ascending node were assigned uniformly and inclinations were drawn from a Rayleigh distribution with $\sigma_i=0.08\approx 4.5^\circ$. The blue, orange, and green lines represent the probabilities that exactly one, two, or three planets transit, respectively. The red dotted line represents the probability that two or more planets transit, and the purple dotted line represents the probability that the outer two planets transit. Overall, at high $J_2$ values, the system is more likely to be seen as a single-transiting system and less likely with two or more transiting planets, despite the outer two planets co-transiting more commonly at high $J_2$ values.
    }
    \label{fig:K619TransProb}
\end{figure}

\begin{figure}
\centering
    \includegraphics[width=\linewidth]{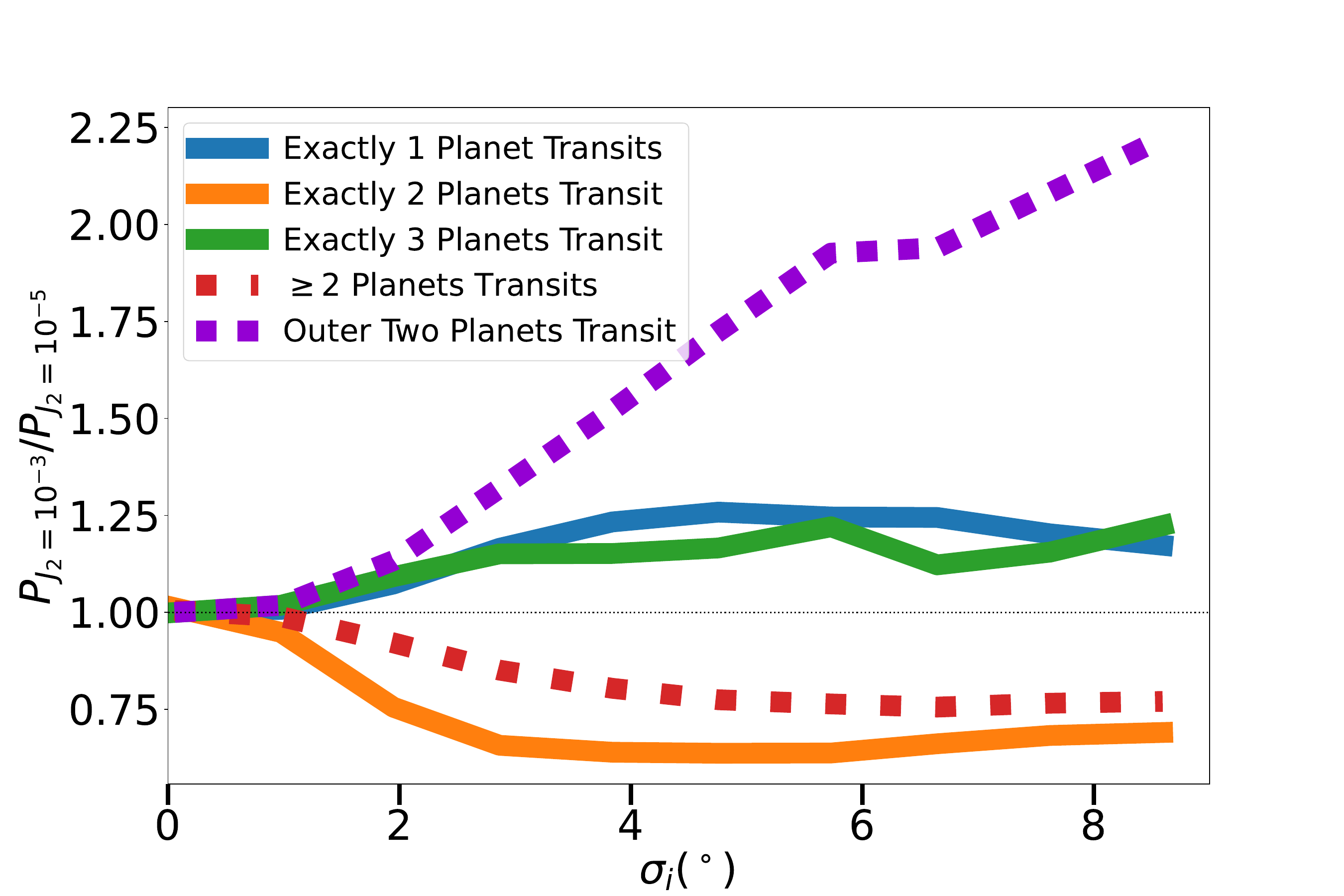}
    \caption{
    Ratios of transit probabilities between initial J2 values of $10^{-3}$ and $10^{-5}$ as a function of the Rayleigh distribution $\sigma_i$ for the initial inclination distribution for Kepler-619. The blue, orange, and green lines represent the probabilities that exactly one, two, or three planets transit, respectively. The red dotted line represents the probability that two or more planets transit, and the purple dotted line represents the probability that the outer two planets transit (irrespective of the inner).
    }
    \label{fig:K619TransProb_sigma}
\end{figure}

\subsection{Transit Probabilities}

As shown in Figure \ref{fig:TimeEvolution}, for Kepler-619 (and more generally), the resonance-crossing aligns the outer two planets of the three-planet system and misaligns the inner relative to them. As discussed, this will have significant effects on the co-transit probabilities. As an example, in Figure \ref{fig:K619OuterPlanets} we show the results of $2000$ Monte-Carlo simulations of Kepler-619 with Rayleigh $\sigma_i = 0.08 \approx 4.5^\circ$. This value of $\sigma_i$ was chosen as it produces median mutual inclinations--namely $\sim 7.4^\circ$, assuming uniformly random longitudes of ascending node for the planets--between planets greater than the observed mutual inclinations between transiting planets. \citet{Zhu+18Architecture} fit a multiplicity-dependent mutual inclination distribution to observed transiting planets and find a median mutual inclination of $\sim 4.8^\circ$.\footnote{The mutual inclination distribution for observed transiting planets is somewhat uncertain. For example, \citet{Fabrycky+14} find that the mode of the mutual inclination distribution of 365 multi-planet systems has a mode between $1^\circ$-$2.2^\circ$. \citet{He+20MaxAMD}, construct a model that asserts exoplanet systems are as close to instability as possible without tipping into it and find that the median mutual inclination between planets in three-planet systems is $2.7^\circ$.} Each simulation's probability of the outer two planets of Kepler-619 transiting together is shown relative to that simulation's initial $J_2$. The dashed purple line corresponds to the mean probability over all initial inclinations at a given $J_2$. When the initial $J_2$ is above the resonance-crossing threshold of $\sim 10^{-4}$, the outer planets align and their co-transit probability is increased by almost a factor of $2$.

During the resonance-induced alignment of the outer two planets, the inner planet misaligns from its companions and remains so. Thanks to its proximity to the star, it is the highest-probability planet to observe in transit (see Figure \ref{fig:TimeEvolution}). Therefore, the probability of seeing all three planets transit together slightly increases due to the sweeping resonance, as seen in Figure \ref{fig:K619TransProb}. However, the large misalignment from its companions reduces the probability of detecting more than one planet in transit, as depicted by the dashed red line.

Figure \ref{fig:K619TransProb} also shows, as functions of initial $J_2$, the mean probabilities that exactly $1$, $2$, $3$, $\geq 2$, and the outer two planets are observed transiting when Kepler-619 is viewed from a random angle for the same $2000$ integrations used in Figure \ref{fig:K619OuterPlanets}. Counterintuitively, the probability that two planets are observed transiting decreases drastically when the initial $J_2$ is high, despite the higher probability that the outer two planets transit together. This is because there are three ways to observe two planets transiting in a three-planet system: inner/middle, inner/outer, and middle/outer. When the outer two align, their combined transit shadow covers a smaller area, and the misalignment of the inner planet reduces its shadow's overlap with that of the outer two. This means that the inner/middle and inner/outer probabilities plummet more than the boost to middle/outer can supplement. Moreover, the probability that any number of planets transit (the $1$ planet transits curve plus the $\geq 2$ planets transit curve) changes relatively little compared to the fractional change in the $1$- or $2$-planet curves. This suggests that these resonances do not significantly alter the probability that the system can be detected in the first place, but rather the multiplicity of that detection.

The change in mean probability of co-transit during a resonance-induced alignment is strongly dependent on the assumed underlying distribution of initial orbital inclinations. Consider a case where every planetary system is born with no mutual inclinations and perfectly circular orbits whose normals are parallel to the stellar spin-axis. Then, as each planet is at its maximum angular momentum for its mass and semimajor axis, the system is dynamically cold and will not be affected if stellar $J_2$ passes through a resonance. Angular momentum would not be transferred between planets, as they each hold the maximum angular momentum permitted by their mass and semimajor axis. Therefore, the underlying inclination distribution is an important factor to consider when evaluating the effects of sweeping $J_2$ resonances on exoplanetary systems\footnote{The case in which all planets are strictly coplanar is strongly disfavored by the evidence. Transiting systems are broadly consistent with having mutual inclinations of a few degrees \citep[see e.g.,][]{Fabrycky+14, chen+22}. If the inclination distribution varies with planetary multiplicity, then larger dispersions are consistent with the observed population \citep[see e.g.,][]{Zhu+18Architecture,He+20MaxAMD,Zhu+21}}. In Figure \ref{fig:K619TransProb_sigma}, we show the results of $20,000$ integrations of Kepler-619's planetary system with varying values of the inclination distribution's Rayleigh $\sigma_i$ ($2,000$ each for $10$ different values of $\sigma_i$). We plot the ratio of several transit probabilities of initial high $J_2$ and initial low $J_2$ as functions of $\sigma_i$. This represents the change in transit probability between the cases in which Kepler-619 was born with a high versus a low $J_2$. As expected, when  $\sigma_i \approx 0\,{\rm rad}$, there is no difference--the transit probabilities are the same regardless of whether the system had high initial $J_2$ or not. However, as the inclination distribution gets warmer, the differences between high and low initial $J_2$ become clear. At higher values of $\sigma_i$, it is expected that the probability that the outer two planets transit together decreases--as they are born with higher mutual inclinations. If a $J_2$ resonance occurs, the outer planets will align such that the warmer the underlying inclination distribution is, the greater the proportional impact of that resonance. Interestingly, the proportional impact on the probabilities that exactly 1, 2, and all 3 planets transit level off after $\sigma_i\sim 3^\circ$. In other words, hot and very hot inclination distributions have the same proportional change in detection probability at high initial $J_2$ compared to low initial $J_2$.

\section{Discussion and Conclusion}\label{sec:conclusion}

We have demonstrated that sweeping resonances, thanks to stellar magnetic braking in a multi-planet system, can significantly affect the transit probability of the system. In initially misaligned systems, angular momentum is transferred from the innermost planet to one or more of the outer companions. This process causes three-planet systems that experience resonance-induced misalignments to split, resulting in an aligned outer pair and a misaligned inner planet.

A natural question is how these resonances affect the observed transiting exoplanet population. Broadly speaking, resonance-induced alignment can increase the transit probability, as illustrated in Figure \ref{fig:TimeEvolution}. For example, in the three-planet system Kepler-619, the resonance-induced alignment increases the transit probability of the outer two planets for initially high $J_2$, i.e., high initial spin. 

Interestingly, through a brief calculation, we can estimate that Kepler-619 had a 50-50 chance of undergoing a sweeping resonance from evolving $J_2$. This estimate is based on Kepler-619's planets' masses and separations, which determine the critical $J_2\sim 10^{-4}$ for the system to experience resonance \citep[e.g.,][see also Figure \ref{fig:K619OuterPlanets}]{Faridani+24}. For a stellar spin distribution adopted from \citet{Gallet+13}, a sun-like star born rotating at a median period of $3.5\,$d will have a $J_2$ at age $10^7$ years of $\sim 10^{-4}$ from Equation (\ref{eq:j2_formula}) for a star of stellar mass and radius, and $k_2=0.28$. This shows that Kepler-619 had an approximately $50\%$ probability of being born spinning fast enough to cause a sweeping resonance in its planetary system. If, like Kepler-619, the typical system as a $50\%$ of being born spinning fast enough, and assuming $20\%$ of systems are susceptible to sweeping resonance \citep[see][]{Faridani+24}, then $\sim 10\%$ of the transiting exoplanet population will have undergone these resonances. Therefore, we suggest that Kepler-619 serves as a good test case for the implications of the sweeping resonance on the configuration and the observational properties. It is worthwhile to note that these resonance-induced misalignments take place at ages of $\sim 10$s of Myr, which is after the disk migration phase and likely after the protoplanetary disks the planets formed from are evaporated \citep[see e.g.,][]{Pfalzner+24+disklifetimes,Ribas+15DiskIsGoneWithin20Myr,Gaidos+25DisksEvaporateBefore20Myr}. 

Specifically, we identify two key factors that affect the resonance behavior and thus the transit probability. These are the initial stellar spin (which determines $J_2$) and the initial misalignment between the planets. Therefore, heuristically, transiting systems fall into one of two outcomes. If the initial misalignment between planets is small \citep[on the order of $<1^\circ$, disfavored by observations, see e.g.,][]{Fabrycky+14,Zhu+18Architecture}, then the resonance will have little effect regardless of the initial stellar spin--as if no resonance occurred at all. Similarly, if the spin is below the critical threshold ($\gtrsim 4$d spin period), no resonance will occur regardless of the misalignments. The second outcome occurs with significant misalignment and high spin. Here, the resonance can occur, and the planets exchange angular momentum (mis)aligning relative to one another. 

For example, as described in Figure \ref{fig:K619OuterPlanets}, assuming a Rayleigh distribution for the mutual inclinations, with $\sigma=4.8^\circ$, an initial high $J_2$ (above $2\times 10^{-4}$ for Kepler-619) has a $75\%$ higher transit probability for the outer two planets compared to initially low $J_2$. In this case, as expected, the probability that only one planet is transiting is also increasing (by $\sim30\%$). Because this gain is mediated by the inner planet misaligning from the outer two, there is a corresponding decrease in the probability that exactly two planets transit (by $\sim 85\%$). The probability that all three planets transit together increases substantially -- by $\sim10$\% -- when the resonance occurs. Overall, this would produce an enhancement in the detection of single-planet transiting systems and a decrease in the number of systems detected as multi-planet. Thus, as highlighted, observing a multi-transiting system in this case is challenging. As closer-orbiting inner planets are more susceptible to these sweeping resonances \citep[i.e., there are more configurations of the outer planets that produce resonance, see][]{Faridani+24}, the misalignment of the inner planet may enhance the mutual inclinations of short- and ultra-short period planets with their companions \citep[as is seen in, e.g.,][]{Dai+18}.

As shown in Figure \ref{fig:K619TransProb_sigma}, the underlying inclination distribution of exoplanets has a major impact on the relative change in transit probability induced by the resonance. For dynamically hotter distributions, the relative impact of the resonance on the transit probability of the outer two planets in a three-planet system is greater. However, for other probabilities (e.g., exactly $1$, $2$, or $3$ planets transiting), the relative impact of the resonance saturates, resulting in warm and hot inclination distributions having the same relative change in detection probabilities. 

Overall, these misalignments should produce an observable signal in the exoplanet statistics. 
This signal would be characterized by reduced numbers of two-planet transiting systems overall. However, we suggest a higher outer two-planet detection rate compared to an estimated intrinsic distribution that does not take resonance-induced (mis-)alignment into account. Specifically, these two transiting planets orbit far enough out from their host star that an undetected inner planet could stably exist.  
Overrepresented detections over predictions of such two-planet systems would indicate that these resonances are more common, even if the inner planets are not detected. As just one example, Kepler-204 hosts two transiting planets with periods of $\sim 14$ and $\sim 26$ days, leaving plenty of space to host an additional planet interior to both that was misaligned from the observed planets by resonance. Moreover, the relative numbers of distant pairs of transiting planets compared to a short-period planet and a nearby companion could constrain the dynamical heat of the underlying inclination distribution. The resonances' relative effect on transit probabilities is dependent on the spread of the inclination distribution. Therefore, the strength of this signal depends on the intrinsic spread of the exoplanet inclination distribution and could constrain it. With NASA's Roman set to observe 100,000 transiting exoplanets \citep[][]{Montet+17RomanPlanets,Wilson+23RomanPlanets}, the sample size for observing this effect will greatly increase.

The reduction in probability for detecting multiple planets caused by the resonance certainly contributes to the effect known as the `Kepler Dichotomy,' where more single-transiting systems were observed by Kepler than predicted \citep[see e.g.,][]{Lissauer+11,Johansen+12, Hansen+13,Ballard+16,Millholland+21}. Misalignments from sweeping $J_2$ resonances represent a way to reproduce some of the dichotomous statistics without requiring separate, distinct distributions to produce the discrepancy. However, as we estimate these resonances occur in $\sim 10\%$ of systems, other channels may also contribute \citep[see, e.g.,][]{Johansen+12,Spalding+16,He+19,MacDonald+20,Millholland+21}.

In higher-multiplicity systems, the picture becomes more complicated. With more planets, there are more opportunities for resonances to occur, and it is not always the innermost planet having its angular momentum transferred away, as a middle planet can be brought into resonance with far-orbiting companions. In this case, any far-orbiting companions, by virtue of their distance from the host star, would very rarely transit compared to the closer planets. By misaligning one of the inner planets relative to the rarely-transiting outer planets, these resonances may also reduce the probability of multiple planets transiting together, similar to what is shown in the three-planet case in Figure \ref{fig:K619TransProb}. Moreover, any significant misalignment of one planet would increase the probability that the system is observed with a single transiting planet, as by definition, misaligning that planet's orbit moves its transit shadow into regions with no overlap with its companions. 

We have found that these resonances should make an imprint on the statistics of transiting planets. By taking into account the required spin for resonance-induced misalignments to occur, we estimated that $10\%$ of three-planet transiting exoplanet systems will have undergone these resonances--a substantial fraction. These misalignments frustrate multiplanet transit detection overall, but enhance the probability of specifically the outermost planets transiting together. This effect could produce a subclass of transiting systems observed as two-planet systems that maintain a hidden, misaligned inner planet within them. Our findings demonstrate a significant, post-migration effect that dramatically impacts the kinds of planetary systems that transit, and how they present themselves.

\section*{Acknowledgments}
 T.H.F. and S.N. thank Howard and Astrid Preston for their generous support. M.R. acknowledges support from Heising-Simons Foundation Grants \#2021-2802 and \#2023-4478.

\bibliographystyle{aasjournal}
\bibliography{kozai, paperexo} 

\bibliographystyle{aasjournal}

\end{document}